\documentstyle[12pt]{article}
\begin{document}

\title{HBT Interference, Wigner functions and MC Simulations}
\author{A.Bialas \\Institute of Physics, Jagellonian
University\\ Reymonta 4,  30-059 Cracow, Poland}
\maketitle
\begin{abstract}
A method of MC simulations including quantum interference, proposed recently by
A.Krzywicki and the present author, is explained.
\end{abstract}

{\bf 1. Introduction.} 

The aim of this talk is to explain details of a scheme
for
MC simulations of multiparticle production including HBT interference
which we proposed recently together with Andrzej Krzywicki \cite{bk}. The
problem became suddenly of great practical importance, when it was realized
that the HBT effects may seriously affect some precise measurements of the standard
model parameters \cite{ls}. As we have heard yesterday from 
Krzysztof Fialkowski, the
existing implementations of quantum interference into standard MC codes suffer
from many problems, theoretical as well as practical ones \cite{h,fw}.
Therefore construction of a viable MC code correcly including the interference
effects is badly needed. To do this, however, it seems first necessary to
formulate the problem and the goals to be achieved. It should perhaps be
emphasized
at this point that the effects of HBT interference are by no means "trivial" or
"automatic", as it is sometimes believed. On the contrary, they depend in
essential way on the physics of the problem. Consequently, there is no
single, unique
method of implementing the HBT interference into existing codes. One must
therefore  be careful to spell out  the underlying physical assumptions.

   We begin in the next section  by a brief reminder of the relation between
the distributions
in x-space and in momentum space which will allow to introduce the necessary
concepts and to formulate the assumptions. In Section 3 the physical meaning of
the procedure is explained in terms of Wigner functions. Some comments and
outlook are given in the last section.

{\bf 2. Density matrix and relation between x-space and p-space.}

   Let
 $\psi (q_1,q_2,...q_N,\alpha)  \equiv \psi(q,\alpha)$ be the probability
amplitude for production of N particles with momenta $q_1,q_2,....q_N \equiv
q$.
$\alpha$ denotes a collection of all other quantum numbers which may be
relevant to the process in question (they may be, e.g., the momenta of other
particles which we do not wish to consider explicitly in a "semi-inclusive"
measurement). The density matrix in momentum space is then
\begin{equation}
\rho(q,q')= \int d\alpha \psi(q,\alpha) \psi^*(q',\alpha)   \label{1}
\end{equation}
This matrix gives all available information about the system in question.
The observed spectrum of particles reads
\begin{equation}
\Omega(q)= \int d\alpha \mid \psi(q,\alpha) \mid ^2  = \rho(q,q)  \label{2}
\end{equation}
We see from this formula that  measurement of the  momentum spectrum provides
only a rather
limited information about the system: only the diagonal elements of the density
matrix are determined.

Let us now consider the
coordinate space. We write
\begin{equation}
\psi(q,\alpha) = \int dx <q \mid x>  \psi(x,\alpha)      \label{3}
\end{equation}
where $\psi(x,\alpha) $ is the  probability amplitude for producing N
particles at the points
($ x_1,x_2,...x_N \equiv x$) and $<q \mid x>$ is the known
transformation
matrix between momentum and coordinate space which shall be specified later.
Introducing the density matrix in coordinate space
\begin{equation}
\rho(x,x') = \int d\alpha \psi(x,\alpha) \psi^*(x',\alpha)  \label{4}
\end{equation}
we obtain the relation
\begin{equation}
\rho(q,q') =\int dx dx' <q\mid x>\rho(x,x') <x'\mid q'>    \label{5}
\end{equation}
which shows that {\it transformation between  description of the system in
momentum
and in coordinate space requires the knowledge of the full density matrix}. The
measured distributions (which give only the diagonal elements) are not enough.

To continue, we need an explicit form of the transformation matrix $<q\mid x>$.
 As it is different for identical and non-identical particles, we shall
treat these two cases separately.

{\bf (i) non-identical particles}

In this case
\begin{equation}
<q\mid x> = exp(iqx) \equiv exp[i(q_1x_1+q_2x_2+...q_Nx_N)]       \label{6}
\end{equation}
(all powers of $2\pi$ are included in normalization of $dx$ and $dq$).
Substituting this into (\ref{5}) we have
\begin{equation}
\rho_0(q,q')= \int dx dx' e^{i(qx-q'x')}\rho(x,x') =
\int dx^+dx^-e^{i(q^-x^++q^+x^-)}\rho(x,x')   \label{7}
\end{equation}
where
\begin{equation}
q^+=\frac12(q+q'); q^- = q-q'; x^+=\frac12(x+x') ; x^- = x-x'    \label{8}
\end{equation}
(from now on we denote the quantities referring to non-identical particles by a
subscript $0$).

From (\ref{7}) we obtain for the spectrum of non-identical particles
\begin{equation}
\Omega_0(q) = \rho_0(q,q) = \int dx^+ dx^- e^{iqx^-} \rho(x^+,x^-).  \label{9}
\end{equation}
This formula shows explicitly that the measured momentum spectrum of
non-identical particles
 does not give any information on distribution of particles in coordinate
space: the $x^+$ dependence is integrated over. Instead, we obtain information
on $x^-$ dependence, i.e., to what degree the density matrix in coordinate
space is non-diagonal. In more physical terms, the momentum
spectrum gives information
only on the {\it coherence properties} of the system in the coordinate space.
Indeed, the off-diagonal part of the density matrix  measures how much
the result of  the  integration over the internal quantum numbers of
the system
(denoted by
$\alpha$ in the Eq.(\ref{4})) is affected by the cancellations due to
"incoherent" summation of terms with "randomly" distributed phases. For
illustration, let us consider a simple parametrizaton of
 the density matrix in the form
\begin{equation}
\rho(x^+,x^-) = \rho(x^+) e^{-(x^-)^2/l_c^2(x^+)}     \label{12}
\end{equation}
where $\rho(x^+)$ is the distribution of particles in coordinate space and
$l_c(x^+)$ is the
"coherence lenght". In this case we obtain
\begin{equation}
\Omega_0(q) = \int dx^+ \rho(x^+) l_c^3(x^+) e^{-4q^2l_c^2(x^+)}      \label{13}
\end{equation}
and we see explicitly that the momentum spectrum measures the average value of
a quantity depending on the coherence lenght $l_c(x^+)$.

 This point is dramatically emphasized for a limiting case of the
system which  is fully incoherent in coordinate space, i.e. for which the
density matrix is purely diagonal ($l_c \rightarrow 0$).
In this case we obtain the spectrum
which is {\it entirely independent} of q! (see e.g. \cite{boal} for a more
detailed discussion of this result and other effects of incoherence).

One final remark about normalization: one sees from (\ref{9}) that
\begin{equation}
\int \Omega_0(q) dq =\int dx^+ \rho(x^+,x^-=0)= \int dx \rho(x,x).
\label{14}
\end{equation}

{\bf (ii) identical particles.}

In this case we have to symmetrize the transformation matrix over
particle momenta and positions and thus we obtain
\begin{equation}
<q\mid x> = \frac1{(N!)^{1/2}} \sum_P e^{iq_Px}   \label{15}
\end{equation}
where P is a permutation of the numbers (1,2,...N) and $q_P$ are the momenta
$(q_1,q_2,...,q_N)$ ordered according to the permutation P. Introducing
(\ref{15}) into (\ref{4}) we have
\begin{equation}
\rho(q,q') =\frac1{N!} \sum_{P,P'} \int dx dx'e^{i(q_Px-q'_{P'}x')}\rho(x,x').
\label{16}
\end{equation}
Using (\ref{7}) this can be rewritten as
\begin{equation}
\rho(q,q')=\frac1{N!} \sum_{P,P'} \rho_0(q_P,q'_{P'})   \label{17}
\end{equation}
so that we obtain for the momentum distribution of the identical particles
\begin{equation}
\Omega(q) = \rho(q,q)=\frac1{N!} \sum_{P,P'} \int dx^+ dx^- e^{i(q_{PP'}^- x^+
+q_{PP'}^+x^-)} \rho(x^+,x^-)   \label{18}
\end{equation}
where $q_{PP'}^+ = \frac12 (q_P+q_{P'})$ and $q_{PP'}^-=q_P-q_{P'}$.
Eq.(\ref{18}) shows explicitly that momentum distribution of identical
particles gives information  on both $x^-$ and $x^+$ dependence of the
density matrix. For the example (\ref{12}) one obtains
\begin{equation}
\Omega(q) =\frac1{N!} \sum_{PP'} \int dx^+ e^{iq_{PP'}^-x^+}\rho(x^+)l_c^3(x^+)
e^{-4 (q_{PP'}^+)^2 l_c^2(x^+)}       \label{19}
\end{equation}
which clearly shows that the dependence on momentum differences $q_{PP'}^-$ is
sensitive to $x^+$ dependence of the particle density in coordinate space
$\rho(x^+)$ and of the "coherence length" $l_c(x^+)$.

Three remarks are in order.

(a) The momentum spectrum of N particles given by (\ref{18}) is expressed in
terms of the density
matrix of N particles in the coordinate space and thus cannot be reduced
(without further assumptions) to the expression involving only single particle
density. In particular, it depends on all N-particle correlations in the
coordinate space. Usually these correlations are neglected (i.e. the density
matrix $\rho(x^+,x^-)$ is written as a product of single particle matrices).
Although
this is a reasonable procedure in the absence of any aditional information,
it should be kept in mind that future data may require to include these
correlations \cite{bk,eg}.

     (b) The normalization of the spectrum (\ref{18}) is different from that of
non-identical particles given by (\ref{14}). Integration over particle momenta
gives
\begin{equation}
\int \Omega(q)dq = \frac1{N!}\sum_{PP'}\int dx_P \rho(x_P,x_{P'})=\int
\Omega_0(q)dq + \sum_{P'\neq P}\int dx_P\rho(x_P,x_{P'}).    \label{20}
\end{equation}
This result shows that the quantum interference changes not only the distribution of 
produced particles but also the production cross-section (i.e. it acts as final-state interaction).

(c) When all particle momenta are equal to each other we obtain from (\ref{2}) and
(\ref{17})

\begin{equation} 
\Omega(q) = N! \Omega_0(q)  \qquad if\  q_1=q_2=...=q_N   \label{20a}
\end{equation}
 consistent with the standard treatment.

{\bf 3.Wigner functions.}

The density matrix has a clear physical meaning, as seen from Eqs. (\ref{1})
and (\ref{4}). Its intuitive meaning is, however, more difficult to grasp.
Therefore it is useful to consider a Fourier transform
\begin{equation}
W(q^+,x^+)= \int dx^- e^{iq^+x^-} \rho(x^+,x^-)   \label {19a}
\end{equation}
which is the generalization of the well known Wigner function (defined usually
for single particle spectrum). It is seen from (\ref{19a}) that $W(q^+,x^+)$ is
a
quantum-mechanical generalization of the classical particle density in momentum
and in coordinate space (Boltzmann phase-space density).

Using (\ref{19a}) and (\ref{9}),(\ref{16}) the particle densities for
non-identical and identical particles can be respectively written as
\begin{equation}
\Omega_0(q) = \int dx W(q,x)   \label{19b}
\end{equation}
\begin{equation}
\Omega(q)=\frac1{N!} \sum_{PP'} \int dx W(\frac{q_P+q_{P'}}2 ,
x)e^{i(q_P-q_{P'})x}
\label{21}
\end{equation}
From these relations one sees again explicitly that while the
momentum distribution of non-identical particles does not give any information
on the particle distribution in x-space, the measured momentum spectrum of
identical
particles is sensitive to x-dependence of the Wigner function, i.e. to
x-dependence of the distribution.

The advantage of using the Wigner functions is that they appeal to one's
intuition (being the analog of the Boltzmann distribution) and thus the
resulting formulas are easier to interpret.
 Of course it should be
kept in mind that this analogy is limited by the fact that a Wigner function
is
-in general- locally not positive
definite. It can oscillate and,
as seen from (\ref{19b}),the oscillations cancel out only after integration over $x$. 
 However, these oscillations can play
a significant role in the Eq.(\ref{21}) for identical particles by
conspiring with oscillating terms in the integrand to contribute significantly
to the result. This is how quantum mechanics shows up in the problem. Thus
regarding Wigner functions as Boltzmann phase-space density is possible only
when they are appropriately smoothed out to remove the oscillations. The price
to pay is that, in general, the resulting probabilistic description can only be
trusted for when the momentum differences in (\ref{21}) are not too large.

{ \bf 4.The proposal for MC simulation.}

Standard Monte Carlo algorithms generate multiparticle events according to an
assumed model for the momentum spectrum $\Omega_0(q)$ which does not include
quantum
interference. The problem is to correct the weights of these Monte Carlo events
once they were generated.

It is clear that this cannot be achieved without additional assumptions. Our
proposal is to assume that
 the corrected spectrum is given by $\Omega(q)$ of the Eq.(\ref{21}) with
{\it the same} Wigner
function as that present in (\ref{19b}).

 I would like to emphasize that this assumption is far from obvious, although
it
is usually accepted without further comments (see, e.g.,\cite{boal}).
It assumes that the
identical and non-identical
particles are produced in the same way. Clearly, this can only be an
approximation (resonance production, for example, influences
differently identical and non-didentical particles).
 As discussed by Bo Andersson at this
meeting,  it is also violated -generally- in the
Lund model \cite{ah}. Hopefully it is not
unreasonable for  events with many particle which we are concerned with
\footnote{To avoid this assumption one needs either a specific model of
multiparicle amplitudes (see e.g. \cite{ah}) or a direct calculation from
the first principles.}.

For an effective use of the Eqs.(\ref{19b},\ref{21}) we need an expression
for the Wigner function which  reproduces the spectrum $\Omega_0(q)$ for
non-identical particles.  Therefore we write
\begin{equation}
W(q,x)=\Omega_0(q) w(q,x).                 \label{22}
\end{equation}
It follows from (\ref{19b}) that $w(q,x)$ obeys the normalization condition
\begin{equation}
\int w(q,x) dx = 1.   \label{22a}
\end{equation}
We see that $w(q,x)$ is the quantum analog of the conditional probability:
given
that particles with momenta $q_1,q_2,...,q_N$ are present in the final
state, $w$ is the probability that they were emitted at the points
$x_1,x_2,...,x_N$.

When (\ref{22}) is inserted into (\ref{21}) we obtain for the correcting
weights
\begin{equation}
S(q) \equiv \frac{\Omega(q)}{\Omega_0(q)} =\frac1{N!} \sum_{PP'}
\frac{\Omega_0(\frac{q_P+q_{P'}}2)}{\Omega_0(q_P)}
\hat{w}(\frac{q_P+q_{P'}}2,q_P-q_{P'})           \label{23}
\end{equation}
where
\begin{equation}
\hat{w}(q,\Delta ) = \int dx w(q,x) e^{i\Delta x}   \label{24}
\end{equation}
with
\begin{equation}
\hat{w}(q,0) = 1.    \label{24a}
\end{equation}

 Clearly, $w(q,x)$ is rather arbitrary
and must eventually be
determined  by  analysis of the data. In absence of any information, and to
exploit fully the intuitive meaning of Wigner functions, we propose -as a
first
step- to neglect possible oscillations and to take $w(q,x)$ in a form which is
everywhere positive definite, so that it can indeed interpreted as a
probability distribution.

The formula (\ref{23}) cannot be used at it stands for most of the existing MC
algorithms because they use an iterative procedure which provides $\Omega_0(q)$
only for one given set of momenta and not all the sets needed in
(\ref{23}). This difficulty can be dealt with by observing that one does not
make a big error by replacing in (\ref{23}) $\Omega_0(\frac{q_P+q_{P'}}2)$ by
$\Omega_0(q_P)$. Indeed, those terms in (\ref{23}) where this approximation is
poor are suppressed by the rapidly decreasing factors $\hat{w}$ and thus need
not be calculated with great precision. Eq.(\ref{23}) now becomes
\begin{equation}
S(q) = \frac1{N!} \sum_{PP'}
\hat{w}(\frac{q_P+q_{P'}}2,q_P-q_{P'})           \label{25}
\end{equation}

The same argument can be used to see that the weights given by (\ref{25}) are positive,
as required for MC simulations. To this end we observe that, as seen from (\ref{24a}), they
are certainly positive if the difference between particle momenta are small. Thus the
positivity is quaranteed in the region where our approximation for Wigner functions is valid.
As we have  argued before, outside of this  region the  non-diagonal
$w(\frac{q_P+q_{P'}}2, q_P-q_{P'})$
are  small and -whether positive or not- do not
play any role in the sum (\ref{25}).

To proceed, further working assumptions are needed. In \cite{bk} we proposed to
start with $w(q,x)$ in a factorized form, each factor being a superposition
of exponentials. Such factorization is most likely not exact (some indications
of this were shown by Hans    Eggers at this meeting \cite{eg}) and may be corrected
when the actual data are fitted. For more details we refer the reader to
\cite{bk}.

{\bf Acknowledgments}
      I would like to thank Wolfram Kittel for  invitation to the meeting and
for a kind hospitality.
 This work has been
supported by
 the KBN grant N0 2 P03B 083 08 and by PECO grant from
the EEC Programme "Human Capital and Mobility", Network "Physics at
High Energy Colliders", Contract Nr: ERBICIPDCT940613.


\begin{thebibliography}{99}
\bibitem{bk}
A.Bialas and A.Krzywicki, Phys.Letters B354 (1995) 134.
\bibitem{ls}
L.Lonnblad and T.Sjostrand, Phys.Letters B351 (1995) 293;
W.J.Metzger, private communication.
\bibitem{h}
S.Haywood, Report Rutherford Lab. RAL 94-07 (1995).
\bibitem{fw}
K.Fialkowski and R.Wit, Z. Phys. C  (in print), and K.Fialkowski, these proceedings.
\bibitem{boal}
D.H.Boal, C.-K.Gelbke and B.K.Jennings, Rev.Mod.Phys. 62 (1990) 553 and references therein.

\bibitem{ah}
B.Andersson and W.Hofmann, Phys.Letters 169B (1986) 364;
B.Andersson, these
proceedings and private communication.
\bibitem{eg}
H.C.Eggers, these proceedings.
\end{thebibliography}
\end{document}